\documentclass[floatfix,reprint,amsmath,amssymb,prapplied,superscriptaddress,aps]{revtex4-2}
\usepackage[dvipsnames]{xcolor}
\usepackage{graphicx}
\usepackage{dcolumn}
\usepackage{bm}

\usepackage{siunitx}
\usepackage{hyperref}
\hypersetup{
colorlinks=true, breaklinks=true, bookmarks=true,bookmarksnumbered,
urlcolor=RoyalBlue, linkcolor=RoyalBlue, citecolor=RoyalBlue, 
pdftitle={},
pdfauthor={\textcopyright}, 
pdfsubject={},
pdfkeywords={}, 
pdfcreator={pdfLaTeX},
pdfproducer={LaTeX}
}

\begin{document}
\title{Skyrmion automotion in confined counter-sensor device geometries}
\author{Kilian Leutner}
\email{kileutne@students.uni-mainz.de}
\affiliation{Johannes Gutenberg University, Mainz, Institute of Physics, Staudinger Weg 7, Germany}
\author{Thomas Brian Winkler}
\affiliation{Johannes Gutenberg University, Mainz, Institute of Physics, Staudinger Weg 7, Germany}
\author{Johannes Güttinger}
\affiliation{Infineon Technologies AG, Villach,
Austria}
\author{Hans Fangohr}
\affiliation{Max-Planck Institute for the Structure and Dynamics of Matter, Luruper Chaussee 149, 22761 Hamburg, Germany}
\affiliation{Faculty of Engineering and Physical Sciences,\\ University of Southampton, SO17 1BJ,\\ Southampton, United Kingdom}
\author{Mathias Kläui}
\email{Klaeui@uni-mainz.de}
\affiliation{Johannes Gutenberg University, Mainz, Institute of Physics, Staudinger Weg 7, Germany}

\date{\today}

\begin{abstract}
    Magnetic skyrmions are topologically stabilized quasi-particles and are promising candidates for energy-efficient applications, such as storage but also logic and sensing. Here we present a new concept for a multi-turn sensor-counter device based on skyrmions, where the number of sensed rotations is encoded in the number of nucleated skyrmions. The skyrmion-boundary force in the confined geometry of the device in combination with the topology-dependent dynamics leads to the effect of automotion for certain geometries. For our case, we describe and investigate this effect with micromagnetic simulations and the coarse-grained Thiele equation in a triangular geometry with an attached reservoir as part of the sensor-counter device.
\end{abstract}

\maketitle

\section{Introduction}
Revolution counters that count the number of rotations are widely used as encoders of components actuated by rotating parts. For safety and reliability, it is desirable that sensing, counting and storing the number of of rotations is realized without an external energy supply. Many safety-relevant applications, such as steering wheel systems, seat belt pretensioner, electromechanical damping systems, clutch actuators, etc., require revolution counting up to tens to hundreds of rotations \cite{diegel1,diegel2}. Currently up to 16 rotations can be sensed based on a magnetic multi-turn counter \cite{novot}. In the required range beyond the 16 rotations, current solutions are prohibitively complex, bulky and expensive, or require active electronic circuits to store the number of rotations, which results in significant stand-by power consumption, re-calibration at service points after power interruptions, and/or regular replacement of environmentally unfriendly and expensive backup batteries. Some suggestions have been put forward to realize magnetic revolution counters beyond 16 rotations \cite{doi:10.1063/1.4728991,PhysRevApplied.8.044004} based on field-induced domain wall motion~\cite{Bisig}. However these approaches have been found to be experimentally challenging, calling for alternative approaches.

\begin{figure*}
    \centering
    \includegraphics[width=\textwidth]{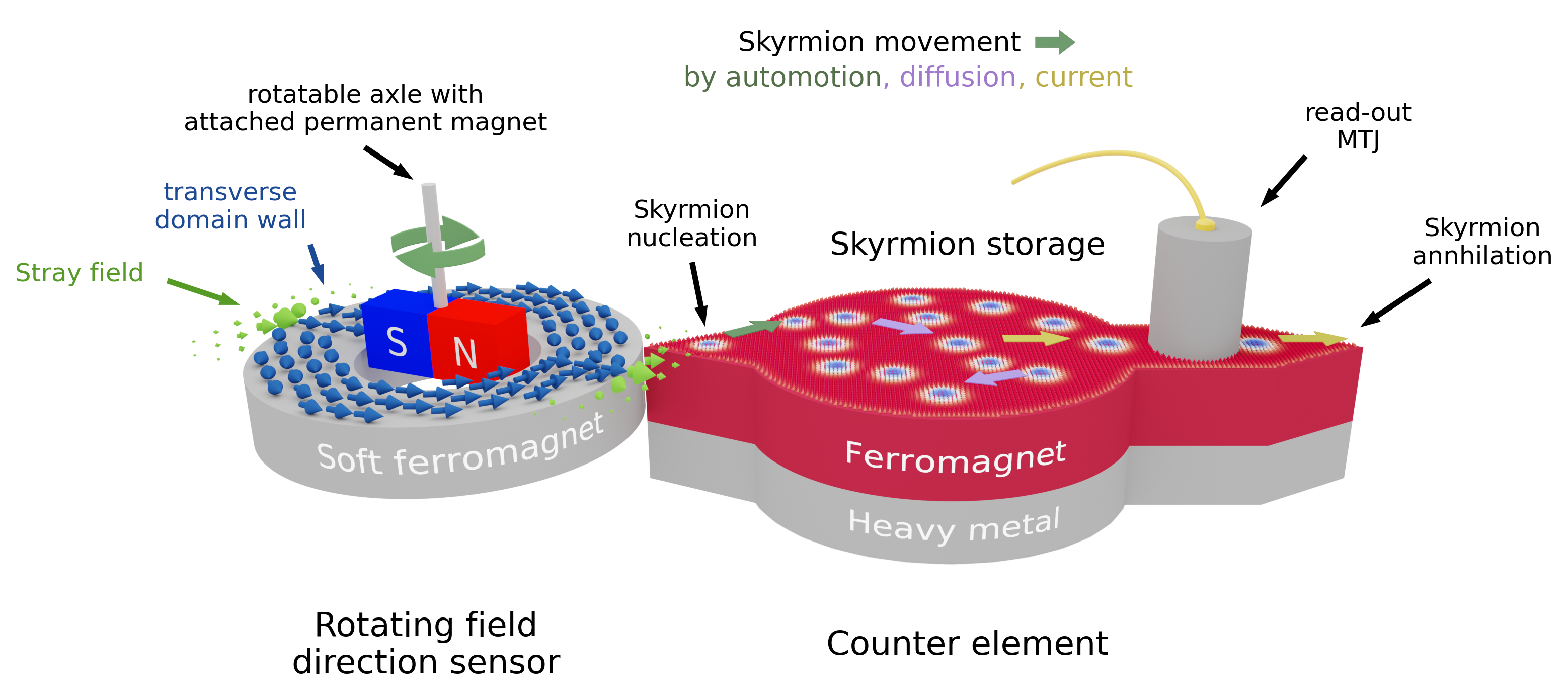}
    \caption{Counter-sensor device comprising the rotating field direction sensor (left) and the counter element (right). Two transverse domain walls in the ring-shaped element of the rotating field direction sensor are moved by the rotation of a permanent magnet which is attached to a rotatable axle. Skyrmions are nucleated in the injector -- in the form of a triangle -- by the sizeable stray field generated by the domain walls. Due to automotion the skyrmions move into the skyrmion storage. Here the skyrmions diffuse randomly. Skyrmions in the storage are moved with a current past a readout element (e.g. MTJ) and are then annihilated at a tip-shaped element. This destructive readout automatically resets the counter to zero at the readout.}
     \label{fig:figure6}
\end{figure*}

A multi-turn sensor-counter concept is depicted in \autoref{fig:figure6}. The rotations of an axle with an attached permanent magnet can be sensed by a rotation field direction sensor which consists of a ring-shaped element with two domain walls. These domain walls produce large magnetic fields \cite{laufenberg}, which therefore can be used for skyrmion nucleation. The skyrmions are nucleated by the in-plane field \cite{zazvorka} and counted in the counter element. The nucleation of skyrmions takes place in a triangular tip and the skyrmions are stored in the skyrmion storage element. From here the skyrmions can be read-out in the readout element and after that annihilated in the skyrmion annihilation element. 

Each revolution of the axle creates a single skyrmion in the triangular tip of the storage element. It is crucial that the skyrmions move out of the nucleation region and into the storage element, see \autoref{fig:figure6}. This movement of skyrmions from the nucleation region into the reservoir can be accomplished with automotion so that during counting no external power source is required. Automotion generally describes the automatic movement of magnetic structures without external energy. The skyrmions move after nucleation via automotion to the large storage element. The number of skyrmions in the storage element thus represents the number of revolutions of the axle and the thermodynamic stability of the skyrmions provides the non-volatility of the counted revolution state. To read out the sensor, a longitudinal current is applied to move the skyrmions past the readout element. Here the skyrmions are detected e.g. with a magnetic tunnel junction element (MTJ). After this, the skyrmions are moved to the annihilation element, where they are removed from the device. Only the readout operation of the sensor requires energy. Unlike existing technologies, this device can count almost arbitrarily large numbers of revolutions since the storage element is scalable.

While the key operations of skyrmion nucleation by in-plane fields \cite{zazvorka}, the motion of domain walls by rotating fields \cite{Bisig} and the current-induced motion and detection of skyrmions has been shown \cite{doi:10.1063/1.5048972}, the automotion of skyrmions is not reported. Automotion has already been demonstrated for domain walls~\cite{PhysRevB.82.214414,doi:10.1063/1.4881061,dwautomotion}, but is not well explored in skyrmionic systems. In our case the automotion of skyrmions in confined geometry is forced by the resulting skyrmion-boundary force and the topology-dependent dynamics. Conventionally skyrmions are moved by spin torques, which are associated with energy consumption \cite{doi:10.1063/1.5048972,skyracetrack,Woo2016}. This is not the case for automotion and therefore appealing for low power applications. 
\section{Skyrmion multi-revolution counter sensor}
\label{sec:counter}
\subsection{Structure of the device} Our novel multi-revolution counter-sensor -- see \autoref{fig:figure6} -- is comprised of a rotating field direction sensor, which senses the direction of an axle with an attached permanent magnet. An annular, ring-shaped element senses the direction of this rotating applied field  by domain wall motion \cite{Bisig}, leading to a localized field that rotates around the structure and an adjacent counter element where for every rotation the stray-field nucleates one single skyrmion, which is then stored, see \autoref{fig:figure6}. By reading out the number of stored skyrmions, the number of revolutions is detected, see \autoref{fig:figure6}. No external power is required during the counting operation, until the result is readout. 
\subsection{Rotating field direction sensor}
The detection of rotations is based on the motion of domain walls in rotating fields as pioneered in Ref.~\citenum{Bisig}. When a $B$-field is rotated, domain walls move in a annular (curved) domain wall conduit structure. For selected geometries, the domain wall is of the transverse wall type, leading to a sizeable stray field \cite{laufenberg}.
\subsection{Counter element}
\subsubsection{Zero-energy nucleation of skyrmions.}
As detailed in \autoref{fig:figure6}, a skyrmion counter element can be positioned adjacent to the annular rotating field direction sensor. It consists of a ferromagnetic layer on a heavy metal layer, so that skyrmions can be stabilized in this material. Every time a domain wall passes by the counter, it locally generates a strong in-plane field at the position of the injector -- in the following in the form of a triangle -- see \autoref{fig:figure6}. Such an in-plane field has been shown to nucleate skyrmions \cite{zazvorka}.
\subsubsection{Zero-energy propagation of skyrmions to the storage element via automotion}
In addition to nucleating a skyrmion by the in-plane stray-field, the skyrmion then needs to move automatically into the storage element. One option is to rely on thermal diffusion \cite{zazvorka,litzius}. However, as this is a stochastic process, a skyrmion might or might not diffuse from the injector area into the storage element before the next rotation count. One would like to have the same magnetic configuration for the injector for every counting event to have the best reliability and reproducibility of the nucleation by the in-plane field. Therefore, a deterministic motion of skyrmions from the injector into the storage element is necessary. To retain the zero external power, we do not want to use currents. We could use a spatial field-gradient who is constant in time. However, this field gradient would involve additional engineering and manufacturing. Instead, we will rely on automotion of skyrmions. By tailoring the shape of the injector for instance in a triangular geometry, the skyrmion will be nucleated at the pointed tip, see \autoref{fig:figure6}, and then move via automotion to the wider part of the triangle to reduce its energy.
\subsubsection{Long-term non-volatile storage of skyrmions in the storage element.}
The skyrmions in the storage element now represent the number of revolutions counted. During storage, skyrmions diffuse in the storage element \cite{zazvorka}. By tuning the materials and geometries we can obtain skyrmion stabilities in the range of $60\,k_B T$ thus being stable for more than $10$ years. The skyrmion storage can be almost arbitrary large and therefore also count almost arbitrary large number of rotations. The reservoir to store the skyrmion needs to be designed to be large enough to host the number of skyrmions that corresponds to the desired number of rotations that are to be sensed, so that no annihilation of skyrmions occurs and diffusion takes place.

\subsubsection{Readout of the storage element.}
To detect the number of revolutions, one needs to read out the number of skyrmions that were generated in the storage element. This can be done by different means. Skyrmions can be detected by a resistance change, via a magneto-transport effect such as a magnetic tunnel junction (MTJ). This resistance signal is detected through CMOS hardware and converted to a machine-readable digital electrical signal. Only from here for the readout, the sensor needs energy consumption. 

\paragraph{Direct readout}One can read out the storage element directly by a magneto-transport effect, as every skyrmion will increase the magnetization in the z-direction and this change can be detected. For example, by a large MTJ on top of the storage element, or, as a fallback option, by the anomalous Hall effect.
\begin{figure*}
    \centering
    \includegraphics[width=\textwidth]{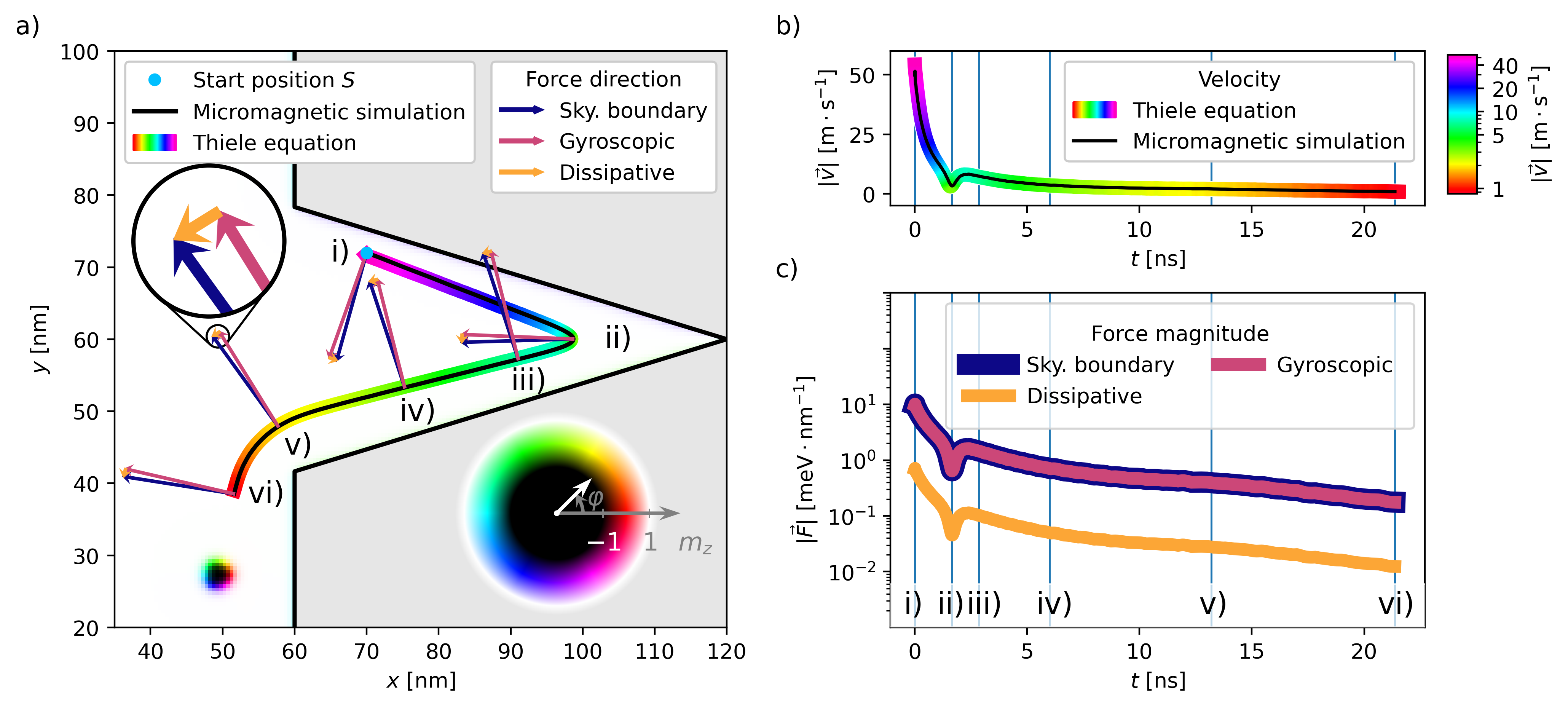}
        \caption{Analysis of the forces in the Thiele equation responsible for the skyrmion dynamics with a opening angle of $\Omega=\SI{34}{\degree}$. The black trajectory is based on the Thiele equation. The colored trajectory is based on micromagnetic simulations, the color indicates the velocity of the skyrmion, the corresponding colorbar is in b). a) Trajectory of a skyrmion nucleated in the upper half of the triangle. For several positions, the direction in the forces of the Thiele equation are depicted. In the background a skyrmion sketch is shown, the description is given by the color wheel.  b) Time dependence of the velocity $|\vec{v}|$ for the trajectory. c)~Magnitude of the forces in the Thiele equation. The labels i) - vi) and the blue lines correspond to the labels in~a). }
     \label{fig:fig1}
\end{figure*}
\paragraph{Readout in a special section}As shown in Ref.~\citenum{litzius}, one can move the skyrmions by spin-orbit torques for instance from the storage element into a readout section, see also indicated readout position in \autoref{fig:figure6}. To realize this, additional electrical contacts are patterned to the counter element where a current flowing will displace the skyrmions in the desired direction of the MTJ readout position. There it can be detected by a     magneto-transport effect such as an MTJ as demonstrated in \cite{penthorn} and indicated as readout MTJ in \autoref{fig:figure6}, or using anomalous Hall effect with additional lateral contacts. This approach uses a pointed tip-shaped element at the end -- indicated in \autoref{fig:figure6} -- to annihilate the skyrmions after readout. This destructive readout mechanism allows one to reset the counter to zero during readout.
\subsubsection{Skyrmion annihilation and reset of the device.}
After the maximum number of revolutions is detected, the storage element needs to be reset (emptied). This can be done in the readout mechanism where the skyrmion can be annihilated after passing past the readout element at a tip-shaped skyrmion annihilation position as previously demonstrated in \cite{zhang}.
For the reset when using readout by detecting the magnetization in the whole storage element, one would apply either a short perpendicular field pulse to annihilate all skyrmions or a short current pulse to heat the system to annihilate the skyrmions.
\section{Skyrmion automotion}
\label{sec:automotion}
We now analyze the missing operation of the skyrmion automotion in detail.
\label{sec:methodmicro}
\subsection{System}
As a material we consider a system hosting small skyrmions, which is biatomic layers of $\mathrm{Pd}/\mathrm{Fe}$ on $\mathrm{Ir}(111)$, the lattice constant of the biatomic layer is 
$a = \SI{0.271}{\nano\meter}$~\cite{Schffer2019}.
The parameters for this material are: saturation magnetization $M_s = \SI{1.1e6}{\ampere\per\meter}$, Dzyaloshinskii-Moriya interaction strength $D = \SI{3.9e-3}{\joule\per\square\meter}$, exchange stiffness $\SI{2e-12}{\joule\per\meter}$, uniaxial anisotropy constant $K_u=\SI{2.5e6}{\joule\per\cubic\meter}$, the uniaxial anisotropy axis $\vec{e}_u = \vec{e}_z$, damping constant $\alpha = \num{0.05}$ and the external field $B = \SI{1.5}{\tesla}\; \vec{e}_z$ \cite{Schffer2019}. The micromagnetic simulations were performed with mumax3~\cite{Vansteenkiste2014,Mulkers2017,Exl2014} at zero temperature. 

We note that we can currently not model much larger systems as experimentally often used~\cite{zazvorka}, but qualitatively the behavior will be the same for both systems. We choose as a sample a thin film with the size  $[\SI{120}{\nano\meter},\,\SI{120}{\nano\meter},\,\SI{0.4}{\nano\meter}]$. For the micromagnetic simulation, the sample is discretized with finite difference mesh and a cell size of $\Delta x = \Delta y = \SI{0.5}{\nano\meter}$, $d = \Delta z = \SI{0.4}{\nano\meter}$.

We define our geometry as follows, see \autoref{fig:fig1}, \autoref{fig:fig2} and \autoref{fig:figsample} in the appendix. The left half ($x \leq \SI{60}{\nano\meter}$) of the sample is completely magnetized and referred as the reservoir.  Further a triangle is spanned by the points $(\SI{120}{\nano\meter},\SI{60}{\nano\meter})$, $(\SI{60}{\nano\meter},\SI{60}{\nano\meter}\pm\SI{60}{\nano\meter} \tan(\Omega/2))$, with $\Omega$ the opening angle at the tip of the triangle, the opposite line of this tip is connected to the reservoir. We define the escape time $t_\text{escape}$ as the time until the skyrmion leaves the triangle
\begin{equation} t_\text{escape}=t(x=\SI{60}{\nano\meter}),
\label{eq:escape}
\end{equation}
that means until the skyrmion has reached a position $x=\SI{60}{\nano\meter}$.
\subsection{Simulation procedure}

In the magnetized area the magnetization is uniform and $\vec{m} = \vec{e}_z$. Then a skyrmion is nucleated in the triangle at the start position $S$. This is achieved by setting the magnetization to $\vec{m}=-\vec{e}_z$ in a circle with the radius $\SI{1.6}{\nano\meter}$ and center $S$. This configuration must be relaxed, but in doing so the skyrmion would move due to the skyrmion-boundary force. Therefore, the magnetization is fixed in the region of a circle with the center $S$ and radius $\SI{0.9}{\nano\meter}$. Then the magnetic configuration is relaxed. To realize this, the Landau-Lifshitz equation is applied without the precession term -- only with the damping term -- until the relaxation process is finished. Then the fixation of the magnetization is removed. The magnetic configuration is then simulated using the Landau-Lifshitz equation, until the skyrmion escapes into the reservoir. Tracking of the skyrmion was performed during the simulations, see \autoref{sec:skytracking} in the appendix for details. 

To compare to the micromagnetic simulations, the skyrmion trajectories were calculated using the Thiele equation \cite{thielepaper}. The skyrmion boundary force was calculated with a micromagnetically computed energy landscape. For details, see \autoref{sec:thiele} in the appendix.
\begin{figure*}
    \centering
    \includegraphics[width=0.75\textwidth]{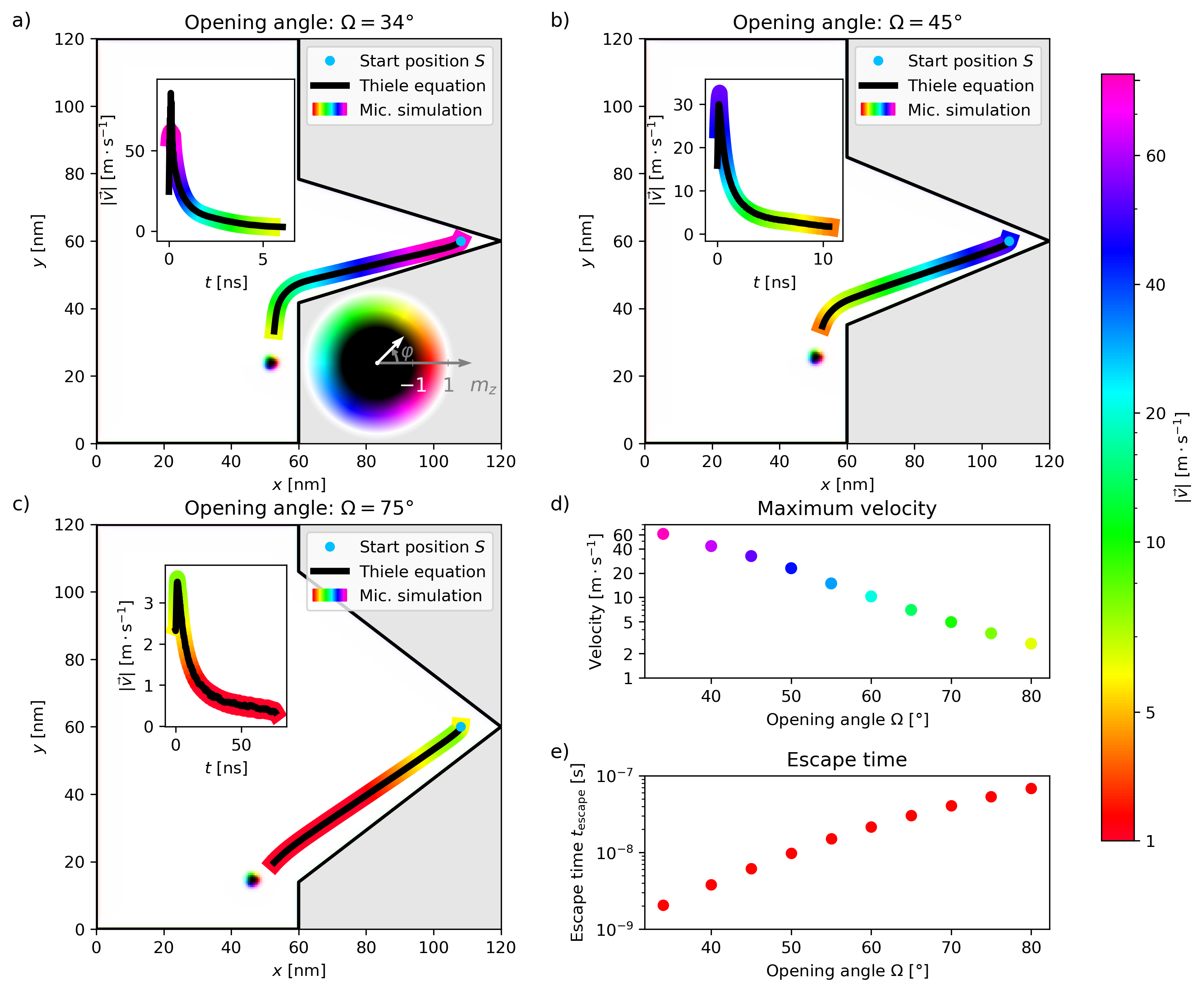}
    \caption{Automotion trajectory of the skyrmion in dependence of the opening angle $\Omega$ in triangle. Same nucleation position for all cases. a),\,b),\,c) depict the dynamics for different opening angles ($\Omega=\SI{34}{\degree}, \SI{45}{\degree}, \SI{75}{\degree}$). The black trajectory is based on the Thiele equation. The colored trajectory is based on micromagnetic simulations, the color indicates the velocity of the skyrmion, colorbar is on the right. Inset depicts the magnitude of the velocity $|\vec{v}|$. In the background a skyrmion sketch is shown, the description is given by the color wheel. d) Maximum velocity decreases with a larger opening angle of the triangle, color of the points corresponds to the colorbar on the right. e) Escape time (see \autoref{eq:escape}) increases with a larger opening angle.}
    \label{fig:fig2}
\end{figure*}
\subsection{Skyrmion automotion modeling results}
\label{subsec:skyrmionautomotionfoundation}
We first study the effective forces on a skyrmion in a Thiele model description (details of the model see \autoref{sec:thiele} in the appendix). As shown in \autoref{fig:fig1}, the dissipative force $\vec{F}_D = D\, \vec{v}$ is perpendicular to the gyroscopic force $\vec{F}_G = G_z \,\vec{e}_z \times \vec{v}$. And as a result from the skyrmion Hall effect, the angle between the skyrmion-boundary force $\vec{F}_\text{sb}$ and dissipative force $\vec{F}_D$ is for our system
\begin{equation}
\theta_\mathrm{SH} = \measuredangle(\vec{F}_D,\vec{F}_\text{sb}) = \arctan\left(\frac{G_z}{D}\right)=  -\ang{86.2}.
\end{equation}
This is the skyrmion Hall angle, which is very large for the small skyrmions considered here. For this material the gyroscopic force is much larger than the dissipative force ($G_z \gg D$), because of the small damping factor $\alpha=\num{0.05}$ and small skyrmion radius $\approx \SI{2}{\nano\meter}$. Due to $\vec{v} \propto \vec{F}_D$ and $\theta_\mathrm{SH}\neq 0$, the automotion dynamics is different from classical mechanics where motion is expected along the gradient of the potential~$\vec{F}_{sb}=-\nabla E$. 

In \autoref{fig:fig1}, we show an example trajectory of skyrmion automotion for the opening angle $\Omega=\SI{34}{\degree}$ and $S=(\SI{70}{\nano\meter},\,\SI{72}{\nano\meter})$. The skyrmion is nucleated in the left upper half of the triangle, see~i) in \autoref{fig:fig1}. Because the skyrmion is close to the upper boundary and away from the lower boundary, one can consider only the upper straight boundary of the triangle. Therefore the skyrmion boundary force points perpendicular to the boundary. Due to the large skyrmion Hall angle, the skyrmion moves almost parallel to the boundary with a small perpendicular component. As the skyrmion moves away from the boundary, the magnitude of the skyrmion boundary force decreases. For our material $G_z \gg D$ is valid, therefore $|\vec{v}| \propto |\vec{F}_\text{sb}(\vec{r})|$, so also the velocity decreases until the skyrmion moves in the vicinity of the triangle apex, see in \autoref{fig:fig1} a) and b) i) to ii).

Conceptually, the skyrmion boundary force in the vicinity of the triangle apex can be thought of as a superposition of two skyrmion boundary forces of two straight boundaries, see \autoref{eq:eqexp} in the appendix. The $\vec{e}_y$ component of the two skyrmion boundary forces cancels out, due to symmetry, the total skyrmion boundary force points in the $-\vec{e}_x$ direction. The dissipative force is oriented by the skyrmion boundary force, in the $-\vec{e}_y$ direction. The skyrmion moves away from the upper boundary towards the lower boundary, see ii) to iii).  Since the $\vec{e}_y$ component of the two skyrmion boundary forces cancels out, the magnitude of the force is lower near the apex. So when the skyrmion moves out of this region towards the lower boundary, the lower skyrmion boundary force predominantly acts, thus the magnitude of the forces and velocity increases and the skyrmion trajectory exhibits a clockwise turn. However, the total energy decreases, due to energy dissipation. During the clockwise rotation of the dissipative force from $-\vec{e}_y$  to the direction parallel to the bottom boundary, the skyrmion moves towards the bottom boundary, see ii) to iii).
\begin{figure*}
    \centering
    \includegraphics[width=.7\textwidth]{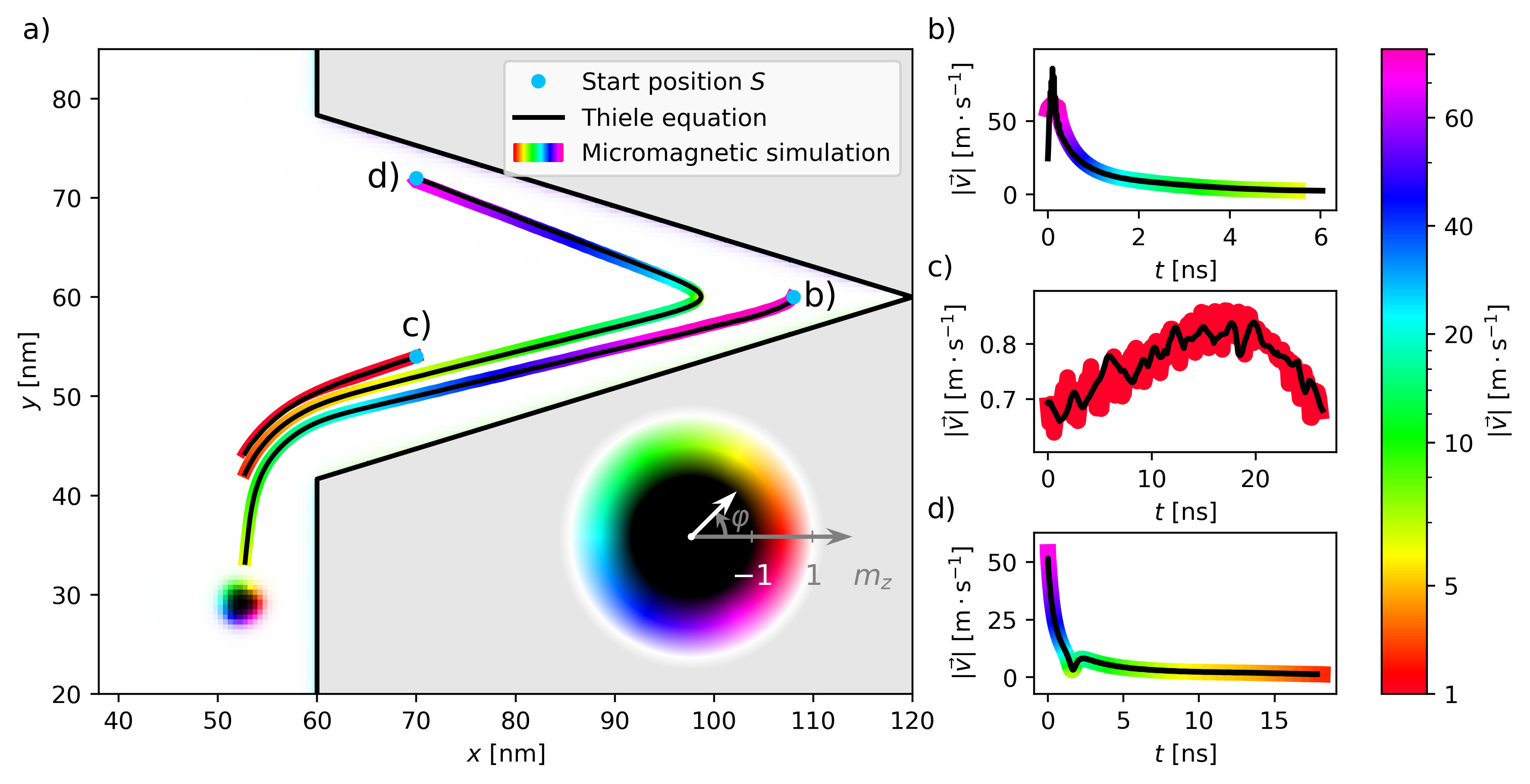}
    \caption{Dynamics of automotion in a triangle ($\Omega=\SI{34}{\degree}$ opening angle) for different starting positions $S$. The black trajectory is based on the Thiele equation. The colored trajectories are based on micromagnetic simulations, the color indicates the velocity of the skyrmion, colorbar is on the right. a) Depicts the different trajectories for different nucleation points. In the background a skyrmion sketch is shown, the description is given by the color wheel. b), c), d) shows the velocity plots for the assigned trajectories in a).} 
    \label{fig:fig3}
\end{figure*}
At position iii) the skyrmion has the minimal distance to the lower boundary. From here on the rotation of the dissipative force continues and the skyrmion moves away from the bottom boundary, see  \autoref{fig:fig1} a) iii) to v). Around position position iii) the upper boundary does not play a significant role for the skyrmion boundary force, therefore the skyrmion boundary force is approximately perpendicular to the lower boundary and the rotation of the forces becomes slower and  vanishes approximately over time. The magnitude of the skyrmion boundary force and the velocity decreases, since the skyrmion moves away from the lower boundary and the upper boundary does not play a predominant role, as seen in  \autoref{fig:fig1} a) and c) iii) to v). \autoref{fig:fig1} c) reveals that the force decreases a bit before iii) and this is due to the fact that we split the skyrmion boundary force in two forces -- for the upper and lower boundary -- for this explanation. The use of \autoref{eq:eqexp} in the appendix is conceptually justified from comparing the exact dynamics of the Thiele equation and the energy landscape calculated with micromagentics.

At \autoref{fig:fig1} a) v) the skyrmion moves to the corner between the triangle and the reservoir and here the skyrmion moves around the corner, due to the skyrmion boundary force, see v) to vi). 

The non-intuitive trajectory is a result of the skyrmion Hall effect, which is expressed by the skyrmion Hall angle. This angle depends also on topology as described by the topological charge $Q=-1$. If $Q=1$, the skyrmion would exhibit a clockwise turn. Every point along a trajectory can be understood as a starting point, due to the fact that Thiele equation is a first order differential equation in time. This also explains -- as long as not indicated differently -- the following presented trajectories for different geometries.
\subsection{Variation of the confined geometry}

\label{subsec:variationgeometry}
We vary the opening angle $\Omega$ of the confined triangle geometry and the results can be seen in \autoref{fig:fig2}. For all angles one obtains qualitatively the same dynamics. Since the nucleation point $S=(\SI{108}{\nano\meter},\SI{60}{\nano\meter})$ is for the different opening angles the same, the force on the skyrmion at the start is smaller for larger angles, due to larger distance from the boundary. Therefore the force curves are smaller, compared to the forces curves for smaller opening angles. Since $|\vec{v}| \propto |\vec{F}_{sb}|$ for $G_z \gg D$, the velocity curves are also smaller for larger opening angles, compared to the velocity curves for smaller opening angles. As shown in \autoref{fig:fig2} d), the maximum velocity decreases with the opening angle, since the skyrmion boundary force decreases with the distance between the skyrmion and the boundary~\mbox{\cite{inverseboltzmann,Schffer2019}.}

\autoref{fig:fig2} e) shows that the escape time (see \autoref{eq:escape}) increases with the opening angle, since  the maximum velocity decreases. The escape time is also larger for larger opening angles because the skyrmion has to travel along the boundary and the boundary length $\SI{60}{\nano\meter}/\cos(\Omega)$ grows for larger opening angles $\Omega$.

For small opening angles $\Omega \le \SI{34}{\degree}$ and nucleation points -- especially at the apex -- too close to the boundary, the skyrmions will move towards the boundary and annihilate instead of escaping into the reservoir.

Good agreement is seen between the Thiele equation with the energy landscape approach and the micromagnetic simulations. This is relevant because it shows that larger skyrmions where the continuous description is even more justified will behave analogously. The small discrepancies between both methods visible in \autoref{fig:fig2} originate mainly from discretization errors and small deviations in the minimization of the system, but also from the fact that the Thiele model does not account for deformation of the skyrmion spin structure.

\subsection{Variation of the nucleation point of the skyrmion}
\begin{figure*}
    \centering
    \includegraphics[width=\textwidth]{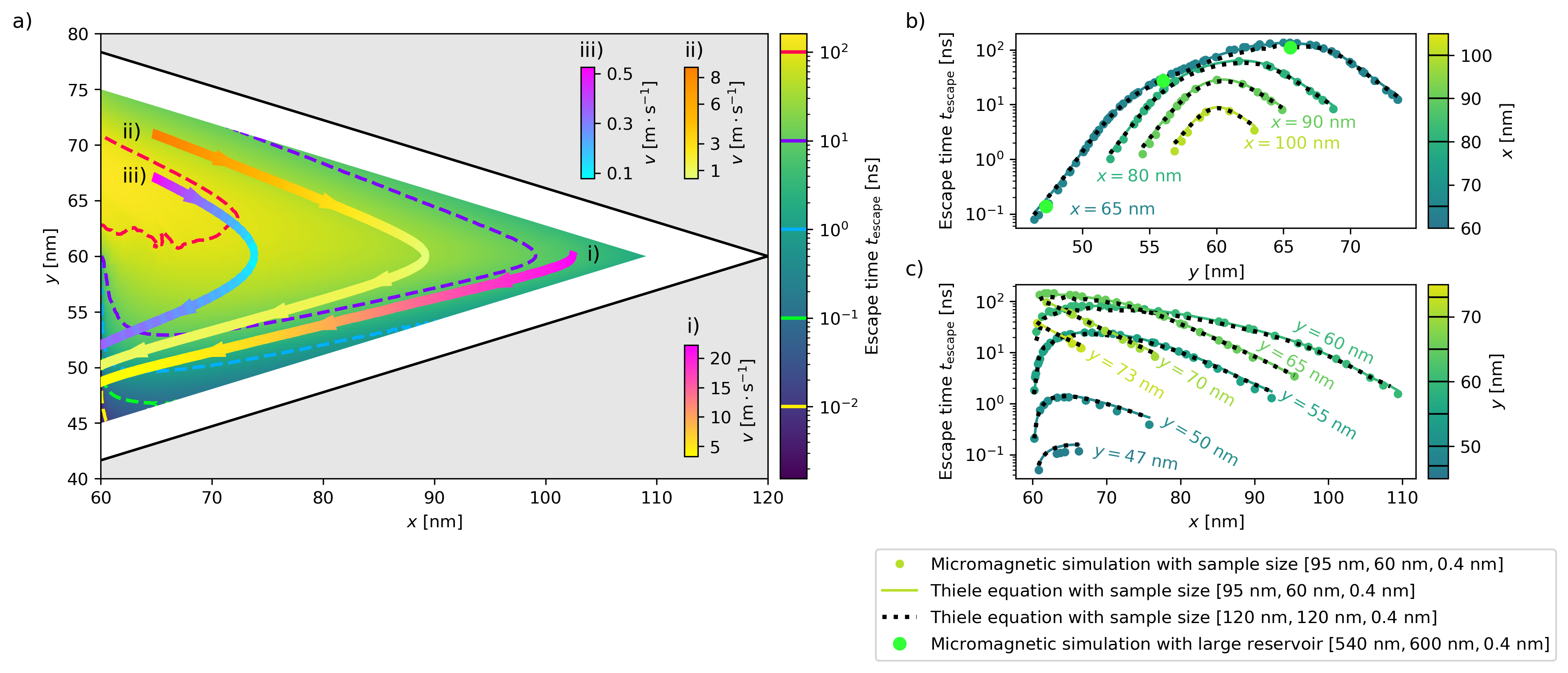}
    \caption{a) Escape time (time for the skyrmion to leave the triangle nucleation region $x<\SI{60}{\nano\meter}$) map calculated with trajectories, based on micromagnetic simulations with a opening angle of $\Omega=\SI{34}{\degree}$. Spatial dependence of the escape time is depicted with the associated colors and contours, colorbar on the right. The three trajectories are based on micromagnetic simulations. The color at each position of the trajectories indicates the velocity, see the three colorbars in the gray area with the assigned label. b) Vertical cuts of the escape time map. Colors indicate the x positions of the cuts. c) Horizontal cuts of the escape time map. Colors indicate the y positions of the cuts. The data in b) and c) for the Thiele equation data points was calculated with the energy landscape approach for the sample sizes $[\SI{95}{\nano\meter},\SI{60}{\nano\meter},\SI{0.4}{\nano\meter}]$ and $[\SI{120}{\nano\meter},\SI{120}{\nano\meter},\SI{0.4}{\nano\meter}]$. The data in b) and c) for the micromagnetic simulation data points is the same as in a) with the sample size $[\SI{95}{\nano\meter},\SI{60}{\nano\meter},\SI{0.4}{\nano\meter}]$, expect for the green data points, here a very large reservoir $[\SI{540}{\nano\meter},\SI{600}{\nano\meter},\SI{0.4}{\nano\meter}]$ with an on the right attached vertical centered triangle with the same dimensions as before is used.}
    \label{fig:fig4}
\end{figure*}

\label{sec:automotionfoundation}
To understand the evolution of the dynamics and the effect of nucleation at different positions, simulations were performed with different starting positions $S$ of the skyrmion as shown in \autoref{fig:fig3} a) for the opening angle $\Omega=\SI{34}{\degree}$. The dynamics for the cases shown in \autoref{fig:fig3} b) and \autoref{fig:fig3} d) are similar to the case discussed in \autoref{subsec:skyrmionautomotionfoundation}. 
The case in \autoref{fig:fig3} c) is however different. The skyrmion moves towards the lower boundary, thus the gradient of the energy landscape increases, but the total energy decreases (if only the lower boundary of the triangle would matter, the gradient would decrease). Shortly after the skyrmion reaches $x=\SI{60}{\nano\meter}$, the velocity decreases, because now the lower boundary is no longer present. The variations in the velocity time dependence for case iii)  -- which are of the order of $\SI{0.05}{\meter\per\second}$ -- are due to discretization errors  and the small velocity $<\SI{1}{\meter\per\second}$. The other velocity-time curves show also variations in the velocity of the same order of magnitude, but these variations are not visible due to a wide velocity range.
\subsection{Escape time map of the confined geometry}
Given the automotion of the skyrmion inside the triangular area, the key question for the device performance is how long it takes for a skyrmion to move to the reservoir. \autoref{fig:fig4} shows the results for this study, where we use \autoref{eq:escape} to define the escape time from the nucleation triangle.
The results are calculated with micromagnetic simulations for the opening angle $\Omega=\SI{34}{\degree}$. For the escape time map, several trajectories were simulated in a smaller sample $[\SI{95}{\nano\meter},\,\SI{60}{\nano\meter},\,\SI{0.4}{\nano\meter}]$ -- where the complete triangle is vertically centered and is on the right side, inside the sample and a corresponding smaller reservoir -- to lower the computational effort. We assign the escape time for each starting point to that position. Each position along a trajectory can be seen as a starting point. The escape time map shown in \autoref{fig:fig4} has been computed from these data using the Clough-Tocher interpolation scheme \cite{2020SciPy-NMeth}.
The Thiele equation with the energy landscape was also used for comparison.

Good agreement is seen between the Thiele equation with the energy landscape approach and the micromagnetic simulation, see \autoref{fig:fig4}. Therefore to rule out that the smaller sample with a smaller reservoir yields an escape time that differs significantly from the sample with the larger reservoir, the Thiele equation was solved also for the large reservoir, with a sample size $[\SI{120}{\nano\meter},\,\SI{120}{\nano\meter},\,\SI{0.4}{\nano\meter}]$. No significant difference is seen between the results in \autoref{fig:fig4}. The relative error between the result of the Thiele equation with the larger reservoir and the micromagnetic simulation is around $15 \%$ and up to $40 \%$ around the boundary of the triangle, but this error is small compared to the significant changes over 3 orders of magnitude and is therefore hardly visible in the plots. The origin of the difference was explained in \autoref{subsec:variationgeometry}.

Two things stand out in particular from the time map. First, the asymmetry in the spatial dependence to leave the triangle nucleation region. In the left upper half of the triangle is the area where automotion lasts longest, due to the left-handed gyroscopic force and the small skyrmion boundary force. From there, clockwise, the escape time decreases. 

We can see that the escape time changes by up to $4$ orders of magnitude. Let us consider the escape time for different y-positions along the line at $x=\SI{65}{\nano\meter}$. Here within $\SI{15}{\nano\meter}$ -- from $\SI{50}{\nano\meter}$ to $\SI{65}{\nano\meter}$ -- the time to leave the triangle changes by $3$ orders of magnitude from $\SI{1e-1}{\nano\second}$ to $\SI{1e2}{\nano\second}$, see \autoref{fig:fig4} a) and b). This is a considerable change of the time scale in a small space, which is a key finding. The main reason behind this is the energy landscape, due to the exponential decrease of energy depending on the distance to the boundary. To check the accuracy and stability of the $3$ order of magnitude in dependence of the reservoir size, a very large reservoir of the size of $[\SI{540}{\nano\meter},\,\SI{600}{\nano\meter},\,\SI{0.4}{\nano\meter}]$ with a vertical centered triangle of the same dimensions as before attached on the right side is used. The resulting green data points in \autoref{fig:fig4}~b) agree well with the rest of the results, the deviation is of the order of magnitude as between the other results mentioned. 

Mechanical rotations occur in the sub-$\si{\mega\hertz}$ speed of revolution regime and result in a skyrmion nucleation frequency up to the sub-$\si{\mega\hertz}$ range. This is also feasible with our concept, since the escape time is smaller than $ \SI{1}{\micro\second}$ for the presented geometry. Beyond detection of mechanical rotation also rotating electrical currents can be detected, for which faster detection can be useful. In this case there is a limit for the detection, depending on the escape time.
\section{Conclusion}
\label{sec:conclusion}
We presented a simple, energy efficient revolution counter sensor concept, which can count an almost arbitrary large numbers of revolutions. Only the readout of the revolutions requires energy consumption. Rotations can also be stored when the skyrmion revolution counter has no power, and can be read out later when power is available.

To realise a device where only the readout requires energy consumption, we rely on demonstrated skyrmion functionality. In addition we use automotion of skyrmions as the transfer mechanism between the nucleation region and the reservoir. Automotion could generally be of interest for energy-saving applications since no external energy needs to be applied.
The skyrmion automotion is also interesting from a physics point of view, due to the interesting non-linear topology-dependent dynamics in the shown case of a triangle. This is expressed in particular by the asymmetric spatial dependence of the escape time and that this time changes over several orders of magnitude in a small region of few nanometers.

The material used, which leads to a large skyrmion Hall angle, is interesting from a conceptual point of view but less suited from an application point of view: the skyrmions move almost parallel to the edge and thus travel a longer distance to get into the reservoir, than they would for a material with a smaller skyrmion Hall angle. 

Geometrical engineering can also be used for rotation direction dependent skyrmion generation. This in turn could allow to distinguish between left- and right-handed rotations.

\section{Acknowledgments}
Thomas Winkler acknowledges funding from the Emergent AI Center, funded by
the Carl-Zeiss-Stiftung, and all authors from Mainz acknowledge the German Research Foundation (DFG SFB TRB 173, SPIN+X, A01 403502522;
SPP Skyrmionics), and the European Research Council
(ERC-2019-SyG, 3D MaGiC, ID: Grant No. 856538). The collaboration between Mainz and Infineon was partly supported by the Austrian Research Promotion Agency (FFG). Financial support for this work also came from the Engineering and Physical Sciences Research Council’s United Kingdom Skyrmion Programme Grant (EP/N032128/1).
\appendix
\section{Confined geometry of the system}
\begin{figure}[h]
    \centering
    \includegraphics[width=0.4\textwidth]{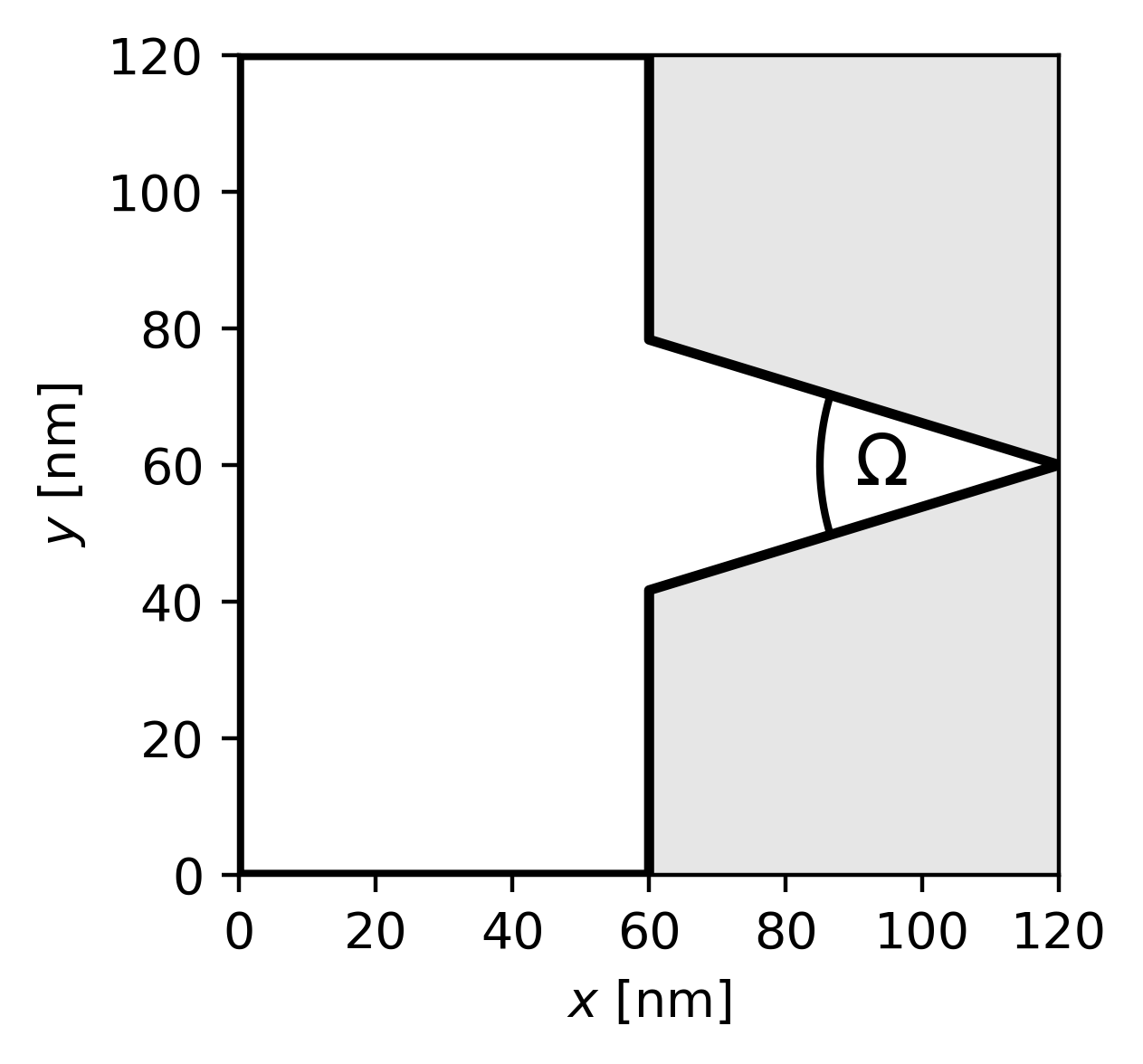}
        \caption{Sketech of the confined geometry of the system. The magnetic region $\vec{m}\neq 0$ is white, non-magnetic region $\vec{m}= 0$ is gray, the black line is the boundary of the sample and $\Omega$ is the opening angle which is $\Omega=\SI{34}{\degree}$ for the sketch.}
     \label{fig:figsample}
\end{figure}
\section{Skyrmion tracking}
\label{sec:skytracking}
During the simulation the skyrmion position $\vec{r}=(x,y)^T$ is tracked.
\begin{equation}
\vec{r} = \frac{\int_A \mathrm{d}^2\vec{r}\;  \frac{1}{2}(1+S_z m_z(\vec{r})) f(m_z(\vec{r}))  \vec{r}}{\int_A \mathrm{d}^2\vec{r}\; \frac{1}{2}(1+S_z m_z(\vec{r})) f(m_z(\vec{r}))} 
\end{equation}
with
\begin{equation}
     f(m_z) = \begin{cases} 1 & S_z = -1\, \wedge\, m_z < -\frac{1}{2}\\1 & S_z = +1\, \wedge\, m_z > +\frac{1}{2} \\0 & \text{otherwise} \end{cases} 
\end{equation}
where $S_z \in \{-1,1\}$ is the sign of $m_z$ in the center of the skyrmion and therefore indicates whether the magnetization in the center of the skyrmion points up or down. The resulting trajectory was smoothed with the Savitzky-Golay filter \cite{2020SciPy-NMeth} to minimize discretization error of the skyrmion position on lengthscales smaller than the grid size. 
\section{Thiele equation}
\label{sec:thiele}

The skyrmion motion can also described by the Thiele equation~\cite{thielepaper,skyrmionintriangle,PhysRevB.87.214419}:
\begin{equation}
\vec{F}(\vec{r}) = \vec{F}_G(\vec{v}(\vec{r})) + \vec{F}_D(\vec{v}(\vec{r})).
\end{equation}
With $\vec{r}$ the position, $\vec{v}(\vec{r})=\dot{\vec{r}}$ the velocity of the skyrmion, $\vec{F}$ the force which acts on the skyrmion, which is equal to the sum of the gyroscopic force $\vec{F}_G$ and dissipative force $\vec{F}_D$. The force which acts on the skyrmion $\vec{F}$ depends on the position $\vec{r}$ and not on the velocity $\vec{v}$, in contrast to the dissipative force $\vec{F}_D$ and gyroscopic force  $\vec{F}_G$ which depend on the velocity $\vec{v}$ and not on the position $\vec{r}$. We are considering a system purely in two dimensions. 
\paragraph{Gyroscopic force}
$\vec{F}_G(\vec{v}(\vec{r})) = G_z \vec{e}_z \times \vec{v}(\vec{r})$ is the gyroscopic force, with the constant 
\begin{equation}
G_z = \frac{M_s d}{\gamma} 4\pi Q \quad \text{with} \quad Q= \frac{1}{4\pi} \int_A \mathrm{d}x\mathrm{d}y\, \vec{m}\cdot \left(\frac{\partial \vec{m}}{\partial x}  \times \frac{\partial \vec{m}}{\partial y} \right).
\end{equation}
Here $d$ is the thickness of the sample (i.e. $d=\SI{0.4}{\nano\meter}$ for our simulations), $\gamma=\frac{g_e e}{2 m_e}$ is the electron gyromagnetic ratio, $Q$ is the topological charge and $A$ the magnetized region ($\vec{m} \neq 0$). The topological charge $Q$ for a skyrmion in our system is $Q=-1$, due to $S_z = -1$ for a skyrmion in our systems \cite{Bttner2018}.
\paragraph{Dissipative force}
$\vec{F}_D(\vec{v}(\vec{r})) = \mathbf{D}\, \vec{v}(\vec{r})$ is the dissipative force, with the tensor
\begin{equation}
D_{ij} = \frac{d M_s}{\gamma} \alpha  \mathcal{K}_{ij}   \quad \text{where} \quad \mathcal{K}_{ij}= \int_A \mathrm{d}x\mathrm{d}y\, \frac{\partial \vec{m}}{\partial x_i} \cdot \frac{\partial \vec{m}}{\partial x_j}.
\end{equation}
For a radial-symmetric skyrmion which we consider in the following: $\mathcal{K}_{xy}=\mathcal{K}_{yx} = 0$, $\mathcal{K}_{xx}=\mathcal{K}_{yy}$ \cite{Bttner2018}. Therefore the dissipative force simplifies to $\vec{F}_D = D\vec{v}=D_{xx} \vec{v}$. To calculate $\mathcal{K}_{xx}$, we  relax a skyrmion with the given material parameters in the center of a sample, and calculate $\mathcal{K}_{xx} = 17.72$.

\paragraph{Force acting on the skyrmion} The force $\vec{F}(\vec{r})$ is minus the gradient of the total energy $\vec{F}=-\nabla E$. The force $\vec{F}$  can be divided into an internal and external force $\vec{F}(\vec{r})=\vec{F}_\text{in}(\vec{r}) + \vec{F}_\text{ex}(\vec{r})$. The reversible external force $\vec{F}_\mathrm{ex}$ is defined as a force due to external applied fields~\cite{thielepaper}. In the work by Thiele \cite{thielepaper} the reversible internal force $\vec{F}_\text{in}$ is zero, due to invariance of the internal energy $E_\text{in}$ with respect to the domain position $\vec{r}$, i.e. $\vec{F}_\text{in} = -\nabla E_\text{in} = 0$. However, this invariance is not given in the case when skyrmion-boundary or skyrmion-skyrmion interaction is taken into account. Considering only one skyrmion the total energy $E$ can be seen as potential energy or energy landscape of the skyrmion and $\vec{F}$ as the Skyrmion boundary force $\vec{F}_\text{sb}$
\begin{equation}
    \vec{F}_\text{sb} = \vec{F} = -\nabla E.
\end{equation}
Note that in Ref.~\citenum{PhysRevB.87.214419} the skyrmion equation of motion was explicit derived with the Thiele approach, here the skyrmion-skyrmion interaction force and skyrmion quenched interaction disorder force appears. Conceptually the skyrmion boundary force can be considered as a force
\begin{equation}
    \vec{F}_\text{sb} = a \exp\left(-\frac{|\vec{t}|}{b}\right) \frac{\vec{t}}{|\vec{t}|} \label{eq:eqexp}
\end{equation} where $a$, $b$ are material dependent constants, $\vec{t}$ is a vector, with the direction perpendicular to the boundary, that starts at the boundary and ends at the skyrmion center. For a straight boundary this result is a good approximation \cite{inverseboltzmann,Schffer2019}. The relative error becomes large for larger distances away from the boundary, due to the approximately flat micromagnetic energy landscape away from the boundary and the exponential dependence of the analytic formula. However the absolute error becomes smaller and smaller.

The approach we use to calculate the two-dimensional energy landscape of a skyrmion in our complex geometry is to place the skyrmion at each point of the sample and fix it in the center. This method was previously used for calculating a one dimensional potential for a straight boundary \cite{Schffer2019}. Then the energy of the magnetic configuration is minimized using the steepest conjugate gradient method. Subsequently, the total energy is calculated. Due to discretization and minimization errors, a bicubic function was fitted over several neighbor cells around one cell and was used for a bicubic interpolation inside the cell. The time integration of the Thiele equation was implemented with the fourth order Runge-Kutta method with adaptive step size.

\bibliography{main}
\end{document}